\documentclass[11pt,twoside]{article}
\usepackage{asp2014}
\usepackage{graphicx,amssymb,sidecap,wrapfig,enumitem}
\usepackage{color}

\aspSuppressVolSlug
\resetcounters

\begin{document}

\title{Science with an ngVLA: Compact binary mergers as traced by gravitational waves}

\author{A.~Corsi$^1$, D.~A.~Frail$^2$, D.~Lazzati$^3$, D.~Carbone$^1$, E.~J.~Murphy$^4$, B.~J.~Owen$^1$, D.~J.~Sand$^5$, R.~O'Shaughnessy$^6$\\
\begin{small}
\affil{$^1$Department of Physics and Astronomy, Texas Tech University, Box 1051, Lubbock, TX 79409, USA; \email{e-mail: alessandra.corsi@ttu.edu}}
\affil{$^2$National Radio Astronomy Observatory, P.O. Box O, Socorro, NM 87801, USA;}
\affil{$^3$Department  of  Physics,  Oregon  State  University, 301  Weniger  Hall,  Corvallis,  OR,  97331,  USA;}
\affil{$^4$National Radio Astronomy Observatory, 520 Edgemont Rd, Charlottesville, VA 22903, USA;}
\affil{$^5$Department of Astronomy and Steward Observatory, University of Arizona, 933 N Cherry Ave, Tucson, AZ 85719, USA.}\end{small}}
\affil{$^6$Center  for  Computational  Relativity  and  Gravitation, Rochester  Institute  of  Technology,  Rochester,  NY  14623, USA.}

\section{Introduction}
Thanks to its  unprecedented sensitivity and resolution, the next generation Very Large Array (ngVLA) has the potential to enable transformational results in the exploration of the dynamic radio sky \citep{BowerMemo}. In light of the recent dazzling discovery of GW170817, a binary neutron star merger whose gravitational wave (GW) chirp \citep{GW170817Discovery} was accompanied by light at all wavelengths \citep[see e.g.][ and references therein]{MMApaper}, here we discuss several new scientific opportunities that would emerge in multi-messenger time-domain astrophysics if a facility like the ngVLA\footnote{Hereafter we assume that the ngVLA will have $\approx 10\times$ the collecting area of the Jansky VLA, operate from 1\,GHz (30\,cm) to 116\,GHz (2.6\,mm) with up to 20 GHz of bandwidth with a compact core for high surface-brightness sensitivity, and extended baselines of at least hundreds of kilometers and ultimately across the continent for high-resolution imaging \citep{BolattoMemo}.} were to work in tandem with ground-based GW detectors:
\begin{enumerate}
\item Probing wide-angle ejecta and off-axis afterglows of neutron star (NS)-NS, and black hole (BH)-NS mergers in low density ISM via Stokes I continuum and polarization measurements;
\item Enabling direct size measurements and dynamical constraints with Very Long baseline interferometry (VLBI) of radio ejecta from NS-NS and/or BH-NS mergers;
\item Unraveling the physics behind the progenitors of BH-BH, BH-NS, and NS-NS mergers via host galaxy studies at radio wavelengths.
\end{enumerate}
In what follows, we briefly describe the scientific landscape expected to be realized when the ngVLA may become operational (Section\,\ref{sec:1}). Then, we discuss radio studies of NS-NS/BH-NS ejecta in Stokes I continuum, linear polarization fraction, and resolved imaging (Sections\,\ref{sec:2} and \ref{sec:3}). We show how the ngVLA could perform resolved host galaxy studies of binary NS-NS and BH-NS mergers (Section\,\ref{sec:4}).  Finally, in Section \,\ref{sec:5}, we summarize and conclude. 

\section{Multi-messenger astronomy in the post-2027 scientific landscape}
\label{sec:1}
During their first two observing runs (O1/O2), the advanced Laser Interferometer Gravitational wave Observatory (LIGO) and Virgo made the big leagues by detecting several BH-BH binaries  \citep[][]{GW151226,GW150914,GW170104,GW170814}  as well as GW170817, a NS-NS merger with an electromagnetic (EM) counterpart \citep{GRB,GW170817Discovery,MMApaper}. Thanks to these discoveries we expect that about 10 years from now, in the ngVLA era, the field of time-domain astronomy will have fully transitioned to time-domain GW astrophysics.  Based on current projections, in the ngVLA era the network of ground-based GW detectors could include Virgo operating at nominal advanced sensitivity, the two advanced LIGO detectors likely in their so-called plus configuration (which foresees a factor of $\sim5$ increase in event rates with respect to advanced LIGO at full sensitivity), the Kamioka Gravitational Wave Detector (KAGRA), and LIGO India. This world-wide network of detectors will be identifying potentially tens to hundreds of GW in-spirals and mergers in the local universe, with localization areas of order $\lesssim 10$\,deg$^2$ \citep[a factor of $\gtrsim 10$ better than today;][]{LivRev}.

As demonstrated by the massive broad-band observational effort that unveiled the EM counterpart of GW170817 \citep[see][and references therein]{MMApaper},  spectroscopic and multi-wavelength observations of GW localization areas are going to be key to finding EM counterparts to GW triggers, removing potential false positives, identifying host galaxies, and constraining several key aspects of the physics of compact binary mergers. Radio and optical observations, for example, crucially complement each other: while optical emission traces the slower thermally-emitting material, radio probes
the non-thermal fastest-moving ejecta. Indeed, in the case of GW170817 \citep{GRB,GW170817Discovery}, optical/IR observations revealed a kilonova and the r-process nucleosynthesis, while the late-time radio and X-ray emissions probed a completely different component, namely, a fast ejecta observed at large angles  \citep[e.g][]{GRB,MMApaper,Coulter2017,Evans2017,Hallinan2017,Kasliwal2017,Lazzati2017,Lazzati2018,Mooley2017,Troja2017,Valenti2017}.

Hereafter, we make the reasonable assumption that by the time the ngVLA becomes operational the community will have collected statistically significant samples of various type of compact binary mergers, and that for at least some of them host galaxies will have been identified \citep[either directly through the detection of EM counterparts, or indirectly through potential anisotropy in their spatial distribution; see e.g.,][and references therein]{Raccanelli2016}.  
It is in this context that the value of a PI-driven radio array such as the ngVLA, with a sensitivity and resolution matched to (or encompassing that of) other post-2027 facilities, is best understood. Here we focus on topics that make use of the superior ngVLA (+VLBI) sensitivity and resolution at frequencies in between those for which SKA1-MID ($\nu\lesssim 1$\,GHz) and ALMA ($\nu\gtrsim 100$\,GHz) will be the premier radio facilities. 

\section{Probing wide-angle ejecta and off-axis jets of NS-NS/BH-NS mergers}
\label{sec:2}
\begin{figure}
\begin{center}
\vbox{
\vspace{-0.8cm}
\includegraphics[height=6.1cm]{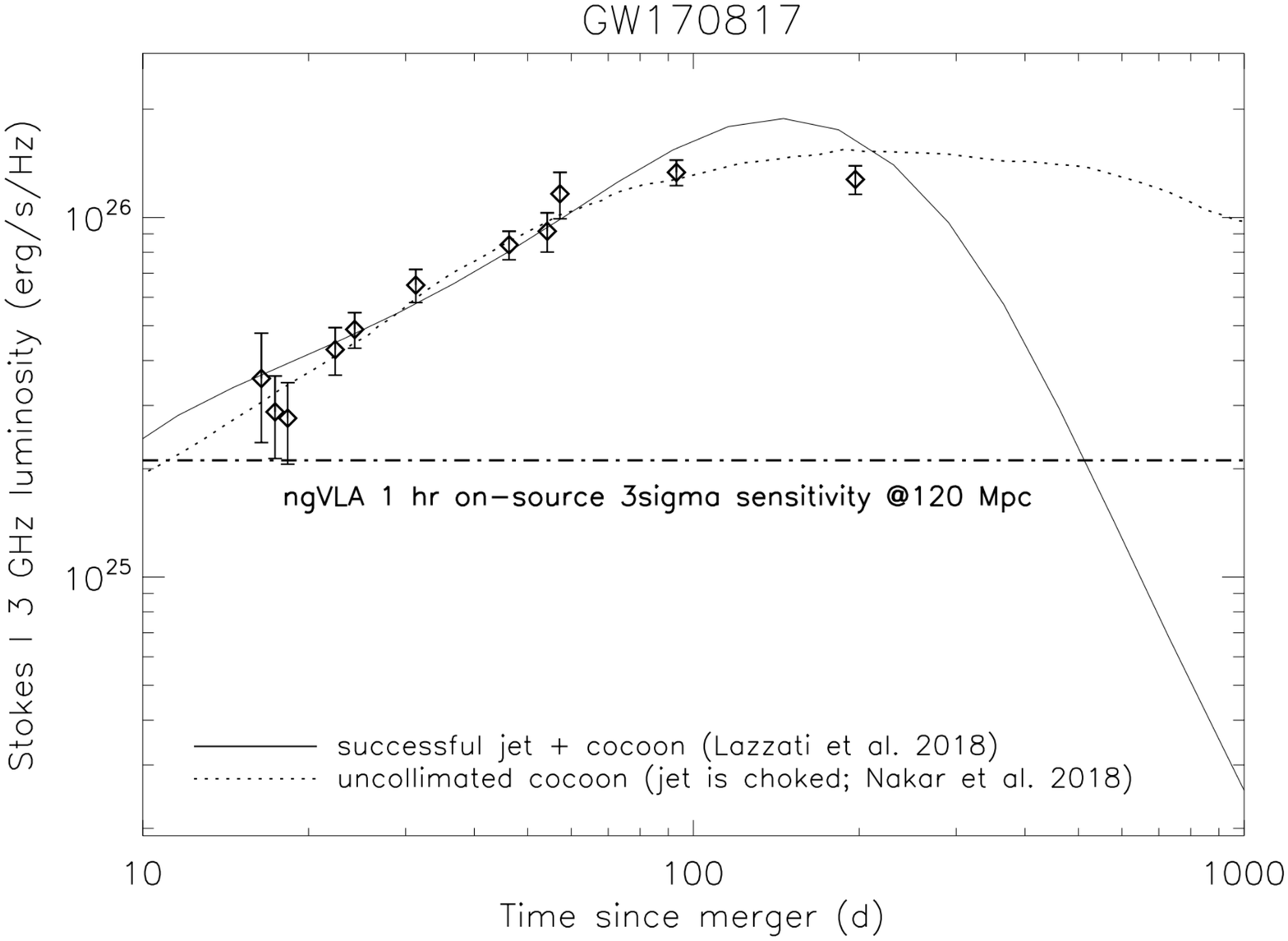}
\vspace{-1.cm}
\includegraphics[height=6.4cm]{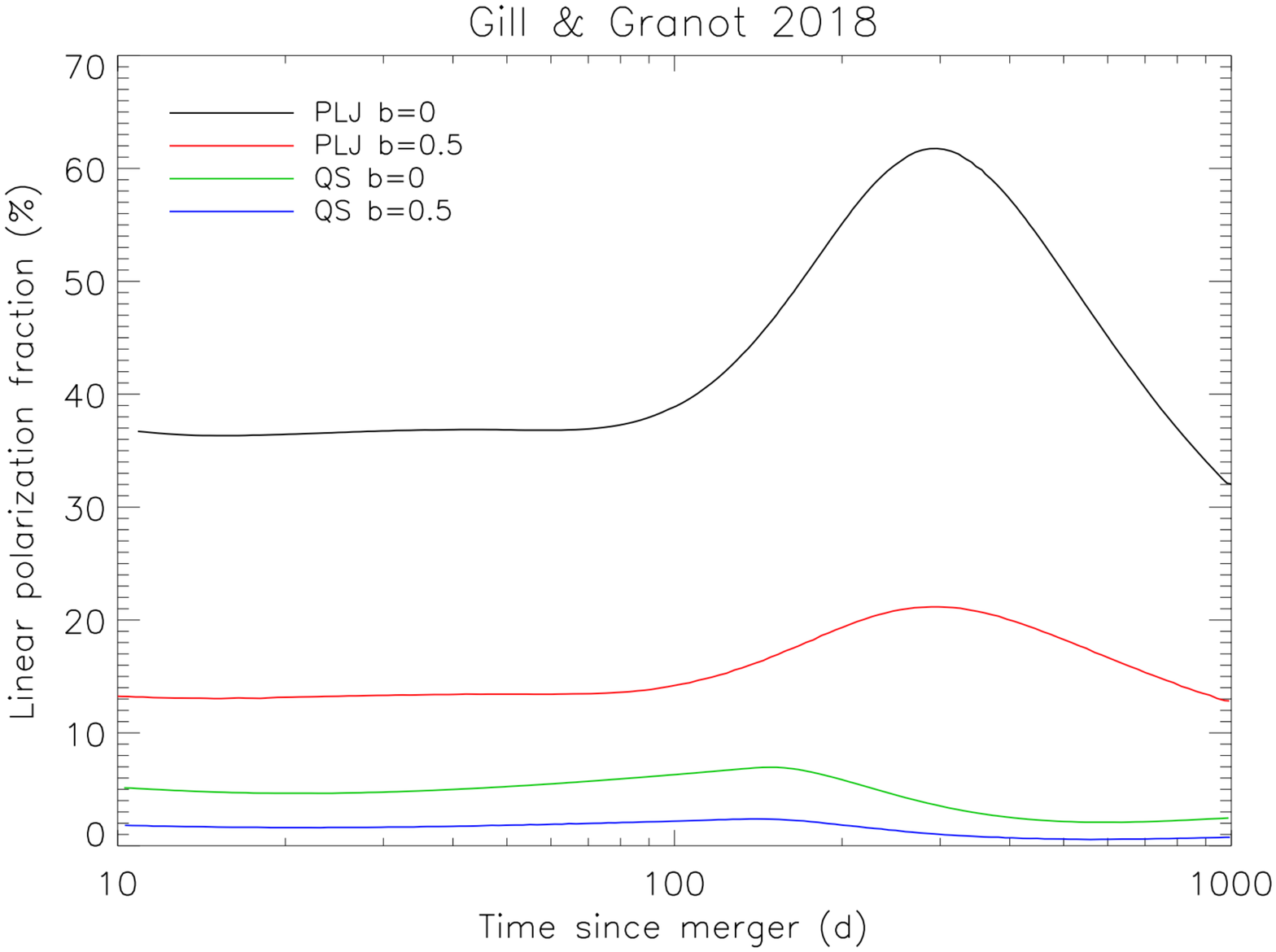}
\vspace{-0.9cm}
\includegraphics[height=6.6cm]{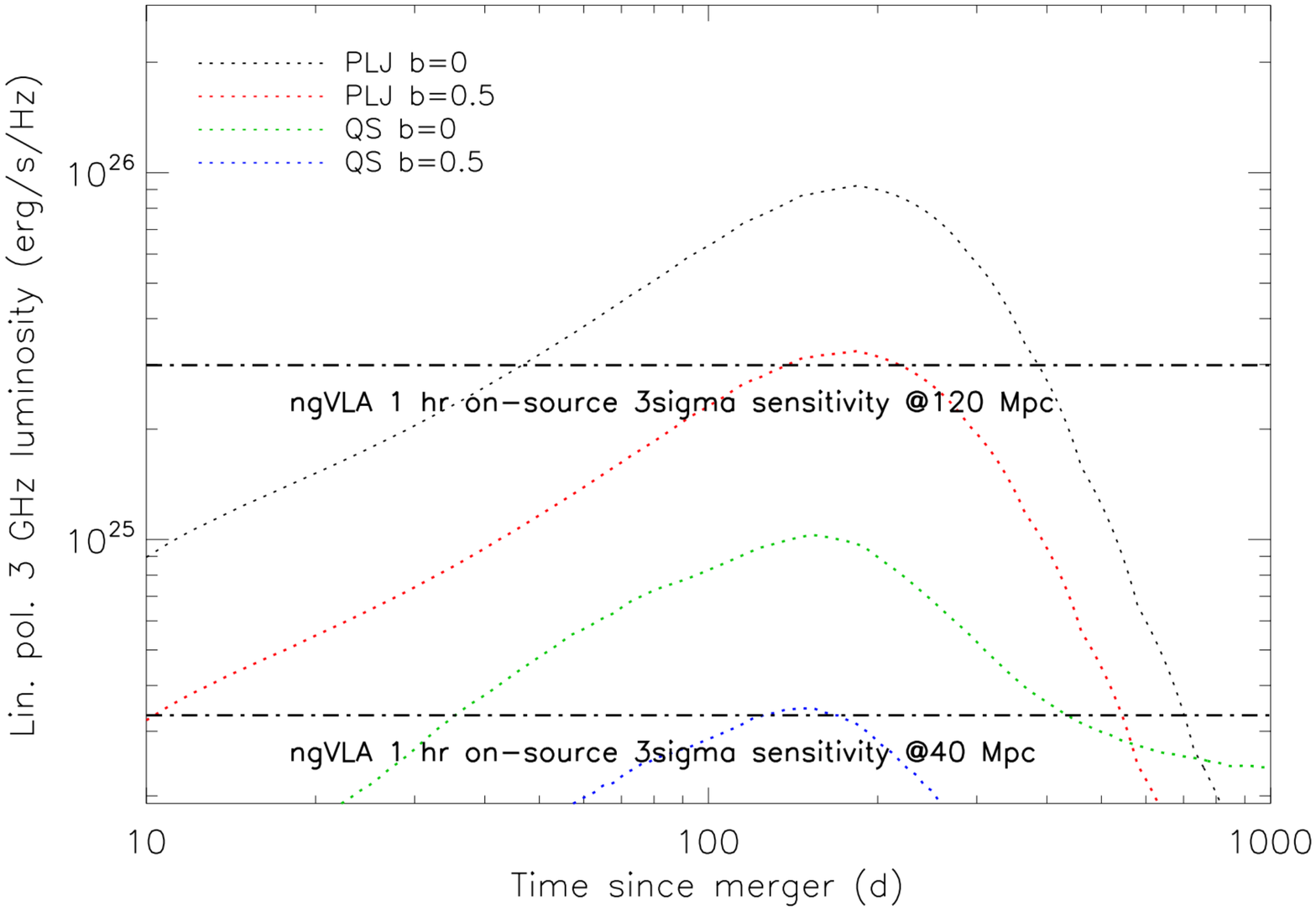}
\caption{\footnotesize{\it TOP: Pre-peak GW170817 3\,GHz light curve \citep[diamonds;][]{Dobie2018} compared with the best fit models in a successful jet plus cocoon scenario \citep[solid line;][]{Lazzati2018}, and in a cocoon with choked jet scenario \citep[dotted line;][]{Nakar2018a}. Only with the ngVLA (rms sensitivity $\approx 0.4\,\mu$Jy in 1\,hr at 2-3\, GHz) we can probe GW170817-like radio counterparts up to the distance horizon for highly significant triple coincidence detections by LIGO and Virgo at their full sensitivty \citep[$d_L\approx 120$\,Mpc, $3\times$ farther than GW170817, for an event rate about $30\times$ higher; see][]{LivRev}. CENTER: Predictions by \citet{Granot2018} for the linear polarization fraction $\sqrt{Q^2+U^2}/I$ of the GHz radio flux for different types of ejecta and magnetic field structures. The quasi-spherical ejecta (QS) best represents a cocoon with a choked jet; the power-law jet (PLJ) best represents the case of a successful jet with a cocoon; the parameter $b$ describes the structure of the magnetic field: for $b=0$ the field is completely in the plane of the shock, and for $b=0.5$ the field component in the direction of the shock normal also contributes. See \citet{Granot2018} for more details. BOTTOM: Best fit Stokes I continuum at 3 GHz from the top panel (solid and dotted lines), multiplied by the linear polarization fraction shown in the center panel for different ejecta and magnetic fields. The ngVLA will probe a large variety of ejecta and magnetic field structures, which are inaccessible to current radio facilities. \label{radiolight}}}}
\end{center}
\end{figure}
Broadly speaking, three major classes of radio counterparts of BH-NS/NS-NS mergers are thought to exist \citep[e.g.,][]{Nakar2011,vanEerten2012,Hotokezaka2016,Lazzati2017,Lazzati2018,Nakar2018b}: (i) Counterparts associated with sub-relativistic merger ejecta producing radio remnants on timescales of a few years; (ii)  ultra-relativistic jets that produce short gamma-ray bursts (GRBs) and (on-axis) radio afterglows in the direction of the jet (evolving on  timescales of a few days); (iii) mildly-relativistic, wide-angle ``cocoons'', also referred to as jet ``wings'', whose emission evolves on timescales of weeks, and which may or may not be accompanied by the contribution of a successful but (initially) off-axis ultra-relativistic jet. A more detailed description of these various components can be found in the contribution to this science book by Nicole Lloyd-Ronning.

While the presence of fast jets that successfully break out from the merger ejecta (scenario (ii) above) is predicted by models positing NS-NS mergers as central engines of short GRBs, the temporal evolution of GW170817 radio afterglow has given evidence for the presence of emission coming from a wide-angle outflow \citep[scenario (iii) above;][]{Kasliwal2017,Hallinan2017,Lazzati2017,Mooley2017,Lazzati2018}. The long-term evolution of GW170817 radio remnant has ultimately confirmed that short GRB-like jet can emerge successfully from NS-NS mergers \citep{Alexander2018,MooleyVLBA,vanEerten2018}, even though when misaligned with our line of sight (off-axis) they lack a bright, fast-decaying afterglow.  Overall, with only one NS-NS merger probed so far, the question of whether successful relativistic jets are formed in \textit{all} binary NS mergers remains open. Answering this question is crucial to understanding whether short GRBs track the NS-NS merger rate, or if instead a larger variety of ejecta outcomes is possible in these mergers, including so-called  ``choked'' jets.  

Building a large sample of NS-NS/BH-NS mergers with radio counterparts is the first step into answering the above question. To this end, we need a radio array like the ngVLA:  as evident from the top panel of Fig. 1, only with the ngVLA we can detect GW170817-like radio counterparts (in Stokes I continuum) up to distances comparable to the horizon for highly significant triple (LIGO+Virgo) coincidence triggers of current ground-based detectors at their full sensitivity \citep[$d_L\approx 120$\,Mpc or $3\times$ farther than GW170817, or an event rate about $30\times$ higher;][]{LivRev}. This conclusion has been confirmed more generally in the recent study by  \citet{Carbone2017}, which shows that even considering a larger variety of possible radio afterglow outcomes from NS-NS mergers (see (i)-(iii) above), the ngVLA will be key to enabling discoveries within a volume large enough for statistically meaningful studies ($\gtrsim 10$ events per year). 
 
While radio continuum monitoring of NS-NS merger ejecta cannot fully remove model degeneracies \citep[see e.g. the top panel of Fig. 1 where the dotted and continuum line largely overlap until a few hundred days since merger;][]{Nakar2018b, Lazzati2018}, radio polarimetry offers the potential of providing a much stronger model discriminant \citep[see Fig. 1 central panel, and e.g. ][ for further discussion]{Rossi2004,Granot2018,Nakar2018a}. Indeed, the sensitivity of an array like the ngVLA will offer us the unprecedented opportunity of probing directly the angular and radial structure of NS-NS merger ejecta, and of their magnetic fields, via radio polarization studies \citep{Corsi2018}.  As shown in the central panel of Fig. 1, a large degree of linear polarization can be considered a smoking gun for the presence of a successful misaligned jet accompanying a large angle outflow in a NS-NS merger.  On the other hand, a quasi-spherical outflow that could result from a choked jet would produce linearly polarized emission at a much lower level. 

As evident from the bottom panel of Fig. 1, we need the sensitivity of the ngVLA to track the temporal evolution of the linear polarization fraction of GW170817-like radio counterparts up to $d_L\sim 120$\,Mpc. The degree of linear polarization can help constrain outflow structure models and, when paired with direct imaging (which can constrain the ejecta structure regardless of the details of magnetic field structure; Section \ref{sec:3}), also enable us to gain critical insight into the possible magnetic field configurations in NS-NS merger ejecta ($b$ parameter in the central panel of Fig. 1). 
\begin{figure}
\begin{center}
\hspace{0.1cm}
\vbox{
\hbox{
\includegraphics[height=4.6cm]{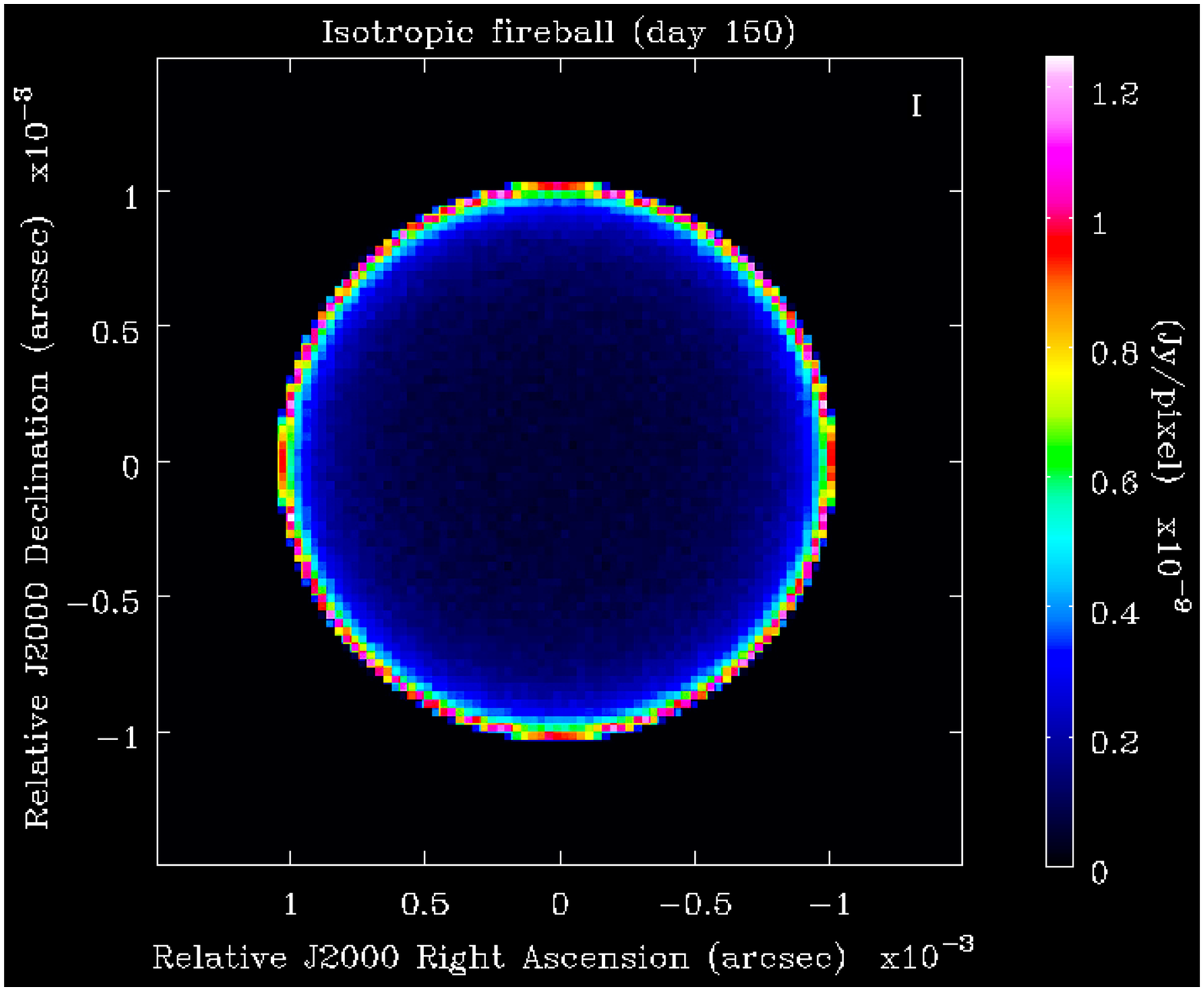}
\hspace{0.05cm}
\includegraphics[height=4.65cm]{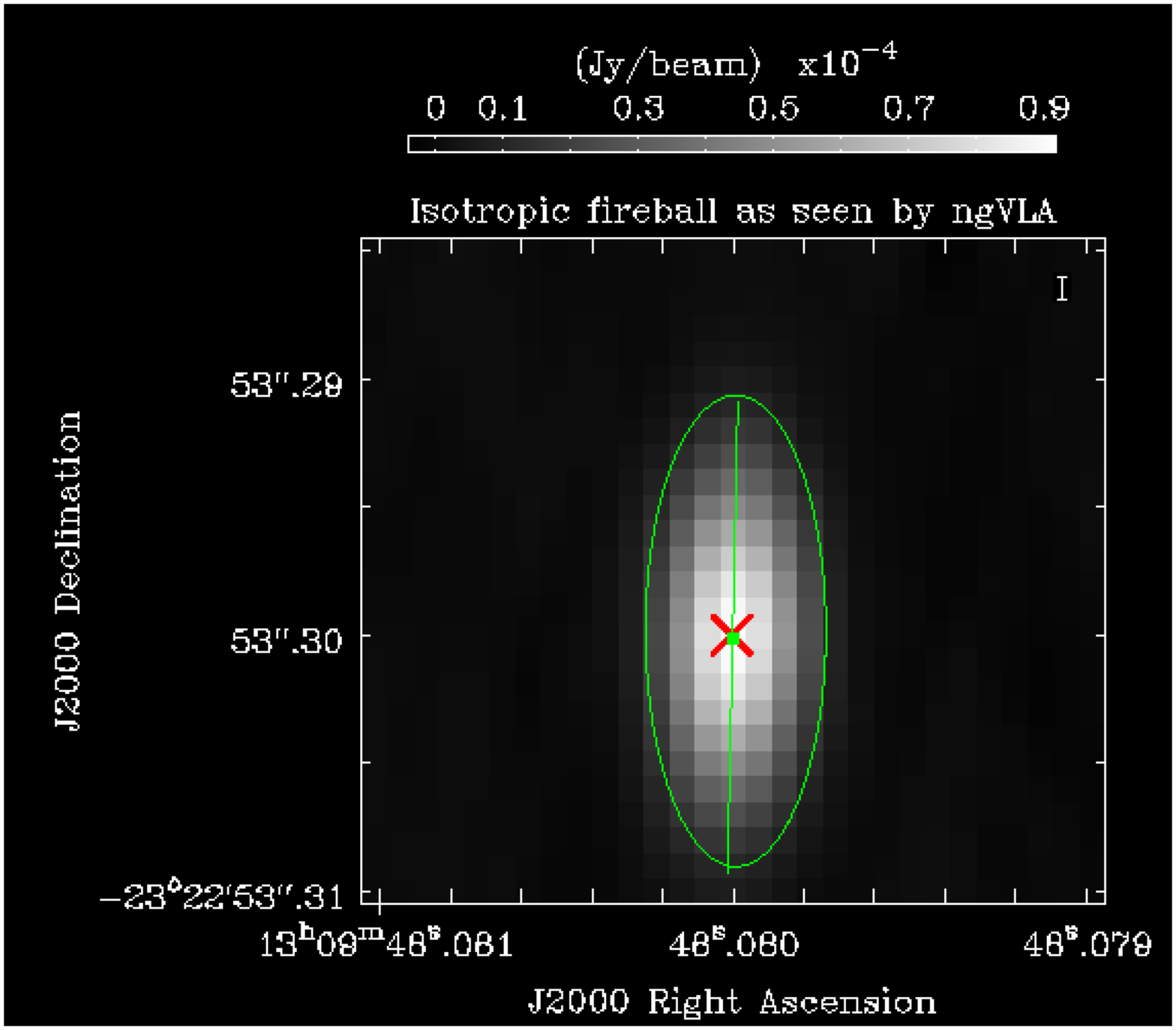}}
\hbox{
\hspace{-0.4cm}
\includegraphics[height=5.1cm]{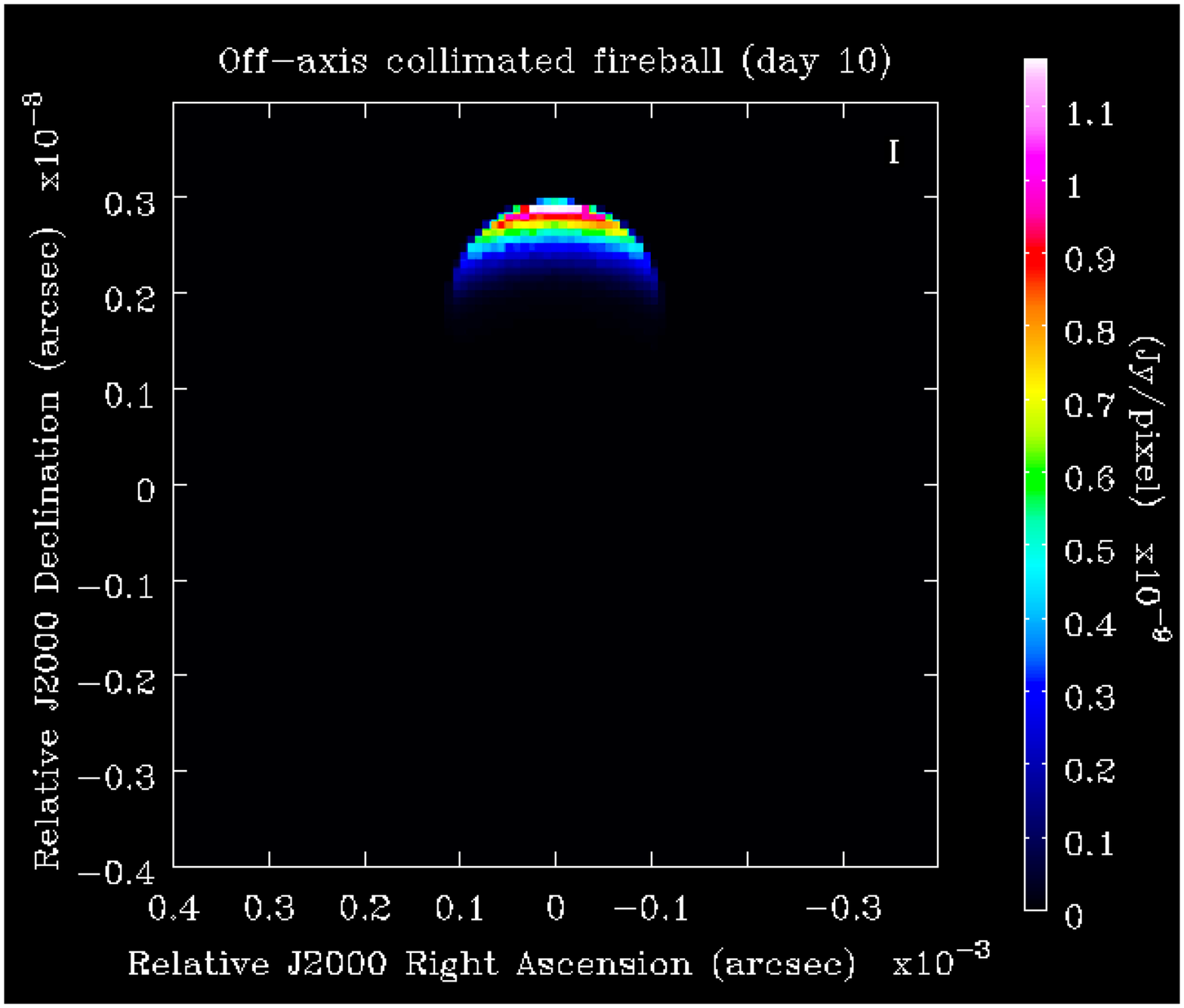}
\hspace{-0.47cm}
\includegraphics[height=5.cm]{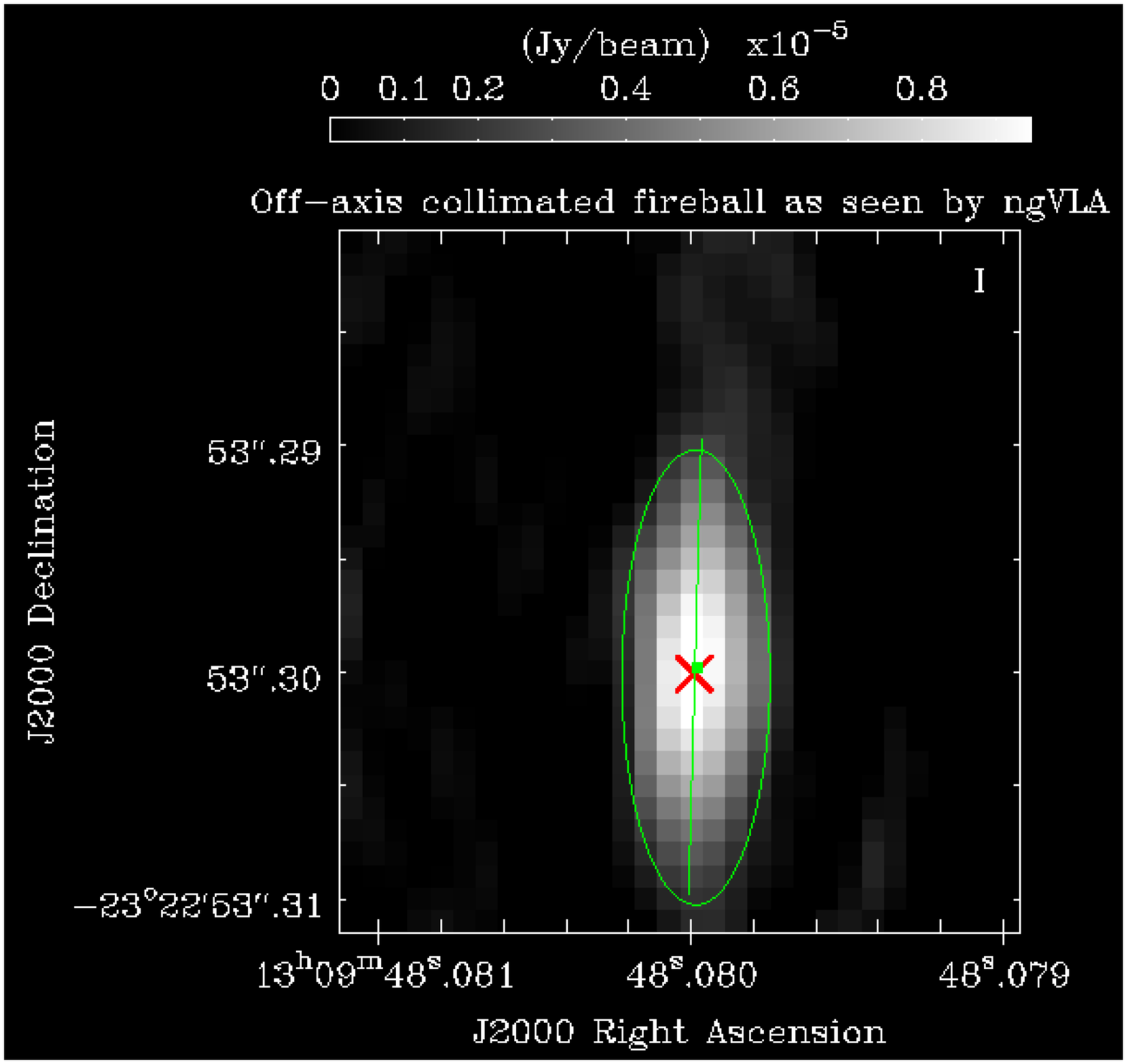}}
\hbox{
\includegraphics[width=6cm]{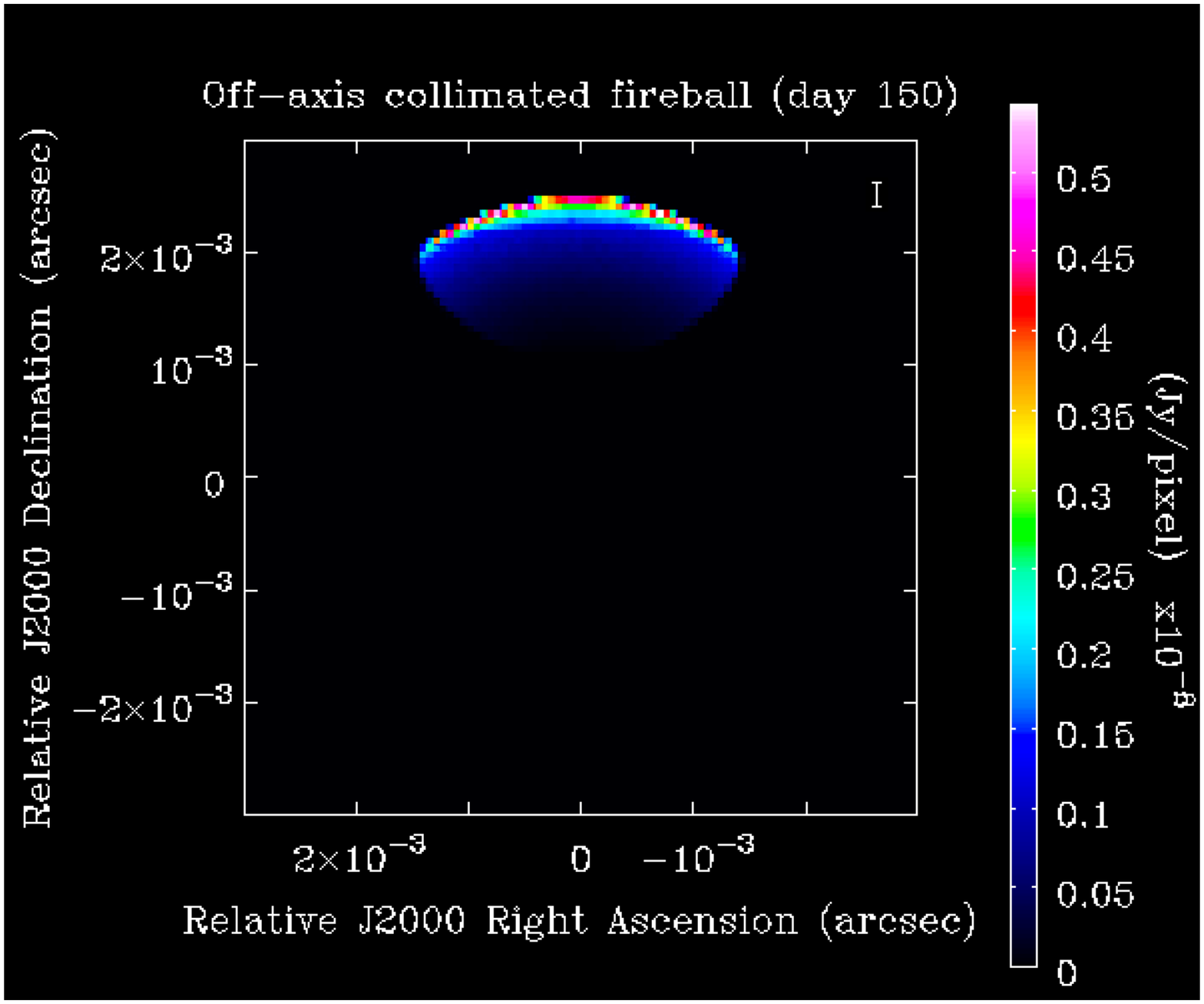}
\hspace{0.02cm}
\includegraphics[height=4.7cm]{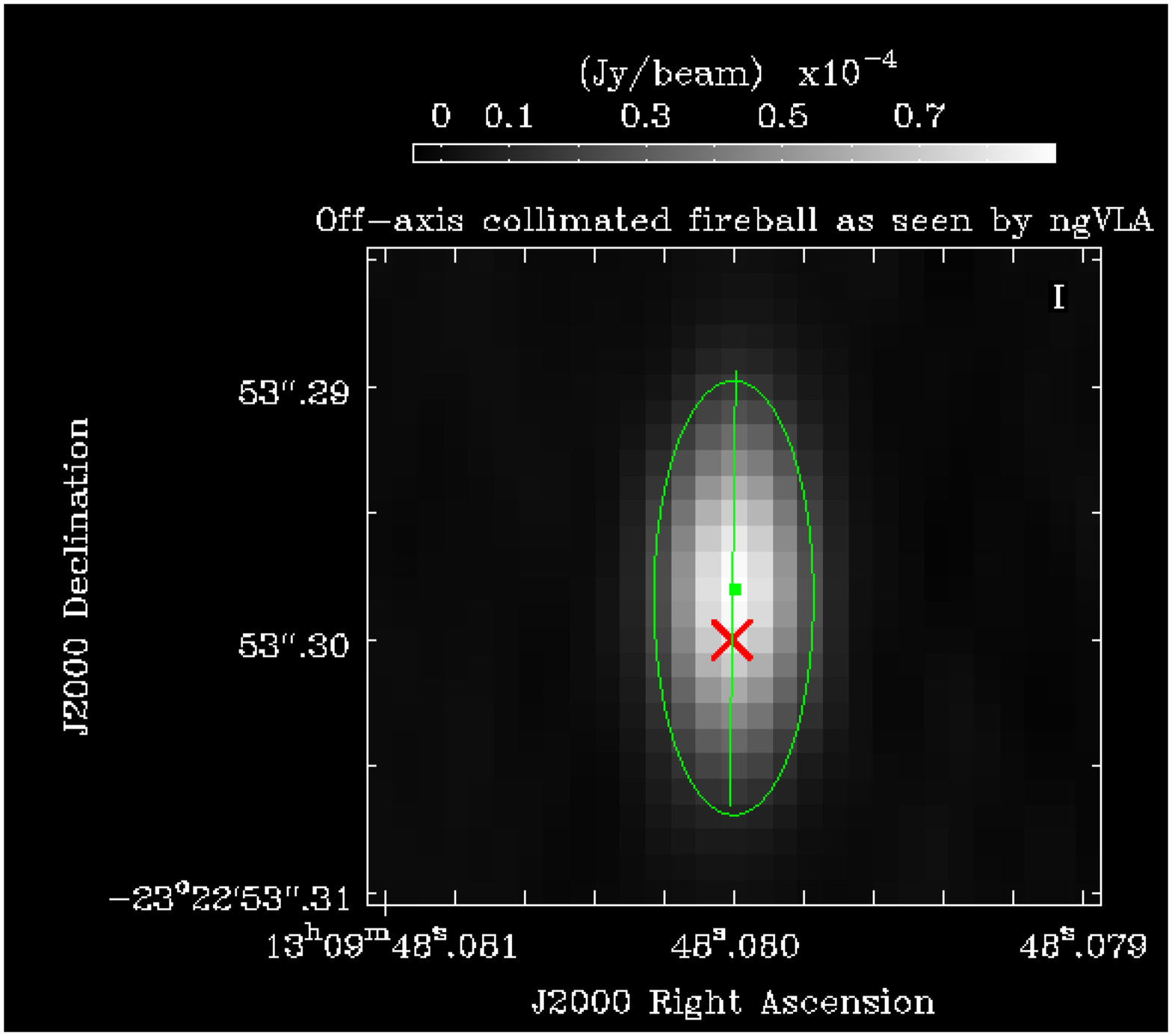}}}
\caption{\footnotesize{\it TOP: Predicted image (left) and simulated observation (right) at 2.4 GHz of an isotropic NS-NS merger ejecta at day $\approx 150$ since merger with isotropic energy $E_{\rm iso}=2.5\times10^{49}$\,erg, fraction of energy in electrons of $\epsilon_e=0.03$, fraction of energy in magnetic fields of $\epsilon_B=0.0002$, expanding in an ISM of density $n_{\rm ISM}=6.8\times10^{-4}$\,cm$^{-3}$ \citep[see][for more details]{Lazzati2018}. CENTER: Predicted image (left) and simulated observation (right) at 2.4 GHz of a top-hat collimated ($\theta_j=23$\,deg) fireball from a NS-NS merger observed off-axis ($\theta_{\rm view}=39$\,deg), at day $\approx 10$ since merger, with $E_{\rm iso}=5.7\times10^{50}$\,erg, $\epsilon_e=0.01$, $\epsilon_B=0.003$, $n_{\rm ISM}=3.2\times10^{-4}$\,cm$^{-3}$. BOTTOM: Same as center but for an observing epoch of 150 d since merger.  In all right panels we have assumed the April 2018 ngVLA-VLBA configuration (see text), 4\,hr integration time at a central frequency of 2.4\,GHz, a nominal 2.3\,GHz bandwidth, and noise rms of $0.2\,\mu$Jy (natural weighting). As evident, the off-axis collimated ejecta would appear to have an emission centroid (green dot) offset from the location of the emission centroid that one would measure at early times and/or e.g. at optical wavelengths (red cross). Moreover, for a collimated off-axis fireball the emission centroid is observed to move over time (compare the location of the green dot in the center- and bottom-right panels relative to the red cross).  \label{offset}}}
\end{center}
\end{figure}

\section{Direct size measurement of radio ejecta from NS-NS and BH-NS mergers}
\label{sec:3}
Resolved imaging of the radio ejecta associated with BH-NS or NS-NS mergers, which requires VLBI techniques, represents the only direct way to map the speed distribution of merger ejecta, and distinguish e.g. collimated relativistic fireballs observed off-axis from quasi-spherical relativistic ejecta. Imaging, coupled with linear polarization studies, could break the degeneracy between outflow and magnetic field structure (see previous section and Fig. 1 central panel). 

The image of a NS-NS merger ejecta will generally depend on the details of the interaction of the fastest ejecta component with with the slower, neutron-rich material, and several different outcomes are possible \citep[see e.g.][]{Nakar2018a}. However, following \citet{Lazzati2018}, here we consider two extreme cases that are likely to roughly ``bracket'' the variety of possible outcomes: an uncollimated (spherical) relativistic ejecta whose image would appear to be ring-like - brighter near the edge and dimmer near the center \citep[see the top-left panel of Fig. 3; see also][]{Granot1999}; and a top-hat (i.e. uniform) collimated relativistic jet whose axis is misaligned with respect to the observer (see the middle- and bottom-left panels of Fig. 2). 

Using the April 2018 updated VLBA configuration of the ngVLA\footnote{This includes the full 214 antennas of the reference array, plus 11 continential-scale antennas: five 18\,m antennas randomly placed at the Green Bank site  and individual 18\,m antennas located at the existing Saint Croix, Hancock, North Liberty, Brewster, Owens Valley, Mauna Kea VLBA sites.}, we have simulated what the ngVLA would see given the emission models shown in the left panels of Fig. 2. We have assumed a 4\,hr-long observation at 2.4\,GHz, and an rms sensitivity of about $0.2\,\mu$Jy/beam (for natural weighting). As shown in the right panels of Fig. 2, an off-axis collimated ejecta would appear to have a time-dependent emission centroid increasingly offset from the location of the counterpart that one would measure at very early times and/or e.g. at optical wavelengths (compare the time-dependent location of the green dot in the middle- and bottom-right panels relative to the fixed red cross). On the other hand, the emission centroid of an isotropic ejecta would be time-invariant and coincide with the optical location at all times (top-right panel in Fig. 2). Moreover, as shown in Fig. 3, the VLBA configuration of the ngVLA will give us enough sensitivity and resolution for mapping directly radio ejecta of NS-NS (and BH-NS) mergers as bright as GW170817 ($\approx 100\,\mu$Jy) around $\approx 150$\,d since merger. Specifically, an analysis of the visibility as a function of baseline will distinguish off-axis collimated outflows from isotropic ones.
\begin{figure}
\begin{center}
\includegraphics[width=\textwidth]{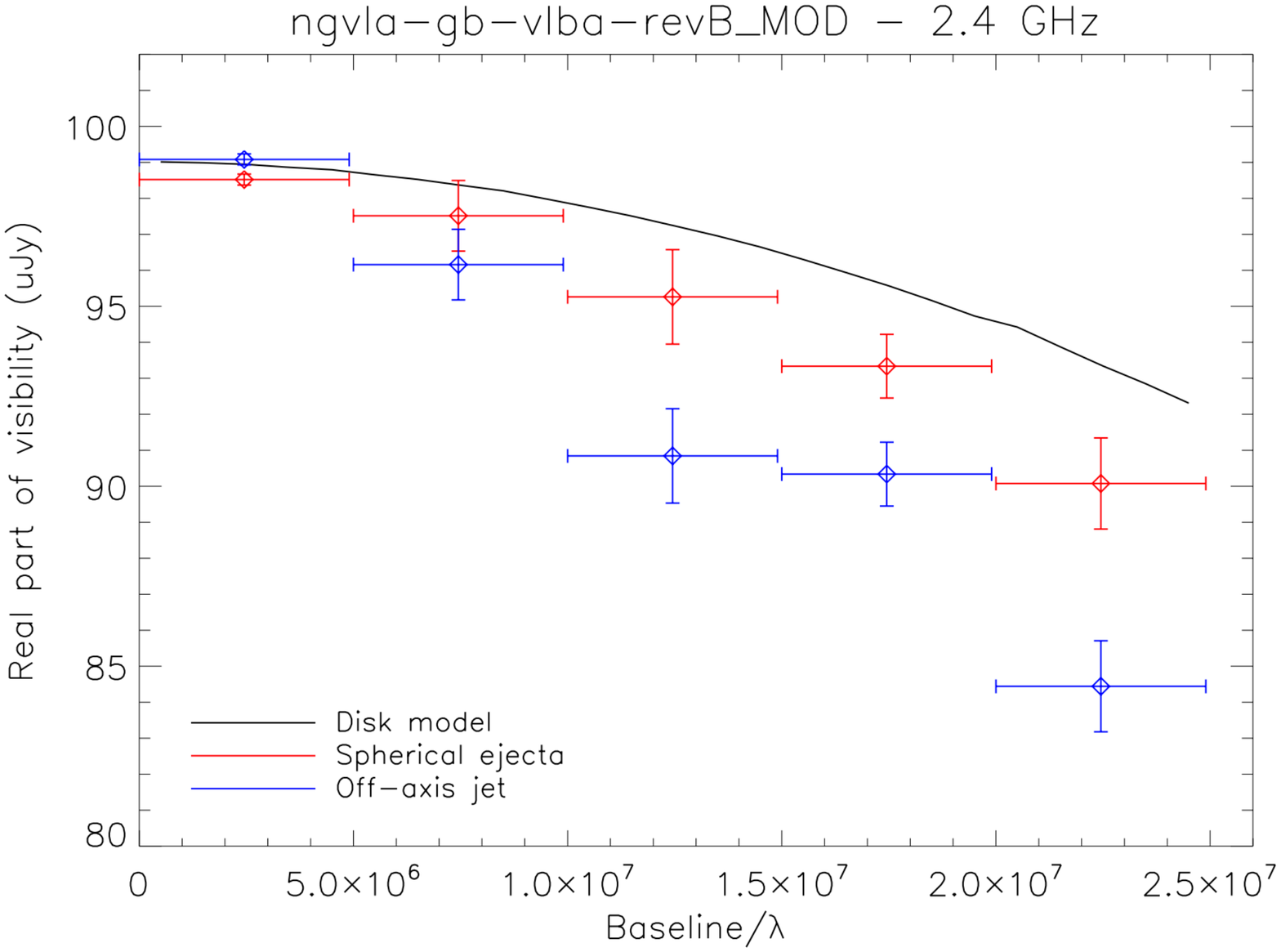}
\vspace{-0.5cm}
\caption{\footnotesize{\it Real part of the ngVLA visibility as a function of baseline for the two models shown in the top and bottom panels of Fig. \ref{offset} (isotropic and off-axis collimated fireballs at at 150 d since explosion when the 2.4\,GHz flux is of order $\approx 100\,\mu$Jy) compared to that of a uniform disk of 2\,mas diameter and total flux of 100$\,\mu$Jy (comparable to the 3\,GHz peak flux of GW170817). We assume the April 2018 ngVLA-VLBA configuration (see text), 4\,hr integration time at a central frequency of 2.4\,GHz, a nominal 2.3\,GHz bandwidth, and noise rms of $0.2\,\mu$Jy (natural weighting). As evident, the ngVLA can resolve and distinguish different ejecta structures. \label{vlba}}}
\end{center}
\end{figure}

\begin{table}
\begin{center}
\begin{footnotesize}
\caption{\footnotesize{\it Host galaxy properties of short GRBs. Form left to right the columns are: GRB name, redshift, host galaxy (optical) SFR, effective host galaxy radius (in optical/IR), expected galaxy radio FWHM at $d_L\approx 120$\,Mpc, 2.4\,GHz luminosity as derived from the optical SFR (see text for more details), 2.4 GHz  radio flux density at $d_L= 120$\,Mpc, and derived radio brightness temperature at 2.4\,GHz and 120\,Mpc (see text for more details).}}
\begin{tabular}{lllll|llll}
\hline
GRB  &   $z$ &  SFR  & $r_{e, z}$ & $r_e$ & $FWHM_{\rm 120\,Mpc}$ & $L_{2.4\,\rm GHz}$ &$F_{2.4\,{\rm GHz, \rm 120\,Mpc}}$ & T$_{2.4\,{\rm GHz}}$\\
          &          & $(M_{\odot}$/yr) & ($''$) &  (kpc) & ($''$) & (erg\,s$^{-1}$\,Hz$^{-1}$) &($\mu$\,Jy) & (K)\\
\hline
061201   & 0.111 & 0.14  &1.09 & 2.2 & 4.0 & $1.5\times10^{27}$ & $8.8\times10$ & 1.2\\
070429B  & 0.902 & 1.1 & 0.65 & 5.1 & 9.2 & $1.2\times10^{28}$ & $6.9\times10^2$ & 1.7\\
070714B  & 0.922  & 0.44 & 0.34 & 2.7 & 4.9 & $4.6\times10^{27}$  & $2.8\times10^2$ & 2.5\\
070724A  & 0.457  & 2.5 & 0.63 & 3.7 & 6.7 & $2.6\times10^{28}$ & $1.6\times10^3$ & 7.5\\
071227   & 0.381 & 0.6 & 0.91 & 4.8 & 8.7 &$6.3\times10^{27}$ & $3.8\times10^2$ & 1.1\\
090510   & 0.903 & 0.3  & 0.93& 7.3 & 13 &$3.1\times10^{27}$ & $1.9\times10^2$ & 0.23 \\
090515   & 0.403  & 0.1  & 1.19 & 6.5 & 12 & $1.0\times10^{27}$  & $6.3\times10$ & 0.097\\
130603B & 0.356 & 1.7 & 0.62 & 3.1 & 5.6 & $1.8\times10^{28}$ & $1.1\times10^3$& 7.3 \\
\hline
\end{tabular}
\end{footnotesize}
\end{center}
\end{table}

\section{Resolving the host galaxies of compact binary mergers}
\label{sec:4}
GW observations of compact object binaries (NS-NS, BH-NS, or BH-BH) can constrain the properties of the compact object themselves, such as masses, spins, and merger rates. However, understanding their progenitors and formation channels (i.e. how do compact binary systems actually form and evolve) requires the identification of host galaxies, and more generally a detailed study of the merger environment. 

A first constraint on the progenitors of compact binary mergers, and their age distribution, would be provided by the demographics of host galaxies. In fact, the distribution of merger timescales impacts the mix of early- and late-type hosts: generally speaking, the smaller the merger delay, the stronger the connection with (recent) star formation (SF) rather than with stellar mass alone, and the larger the late-type fraction \citep[e.g.,][]{Zheng2007}. 
\begin{figure}
\begin{center}
\includegraphics[width=\textwidth]{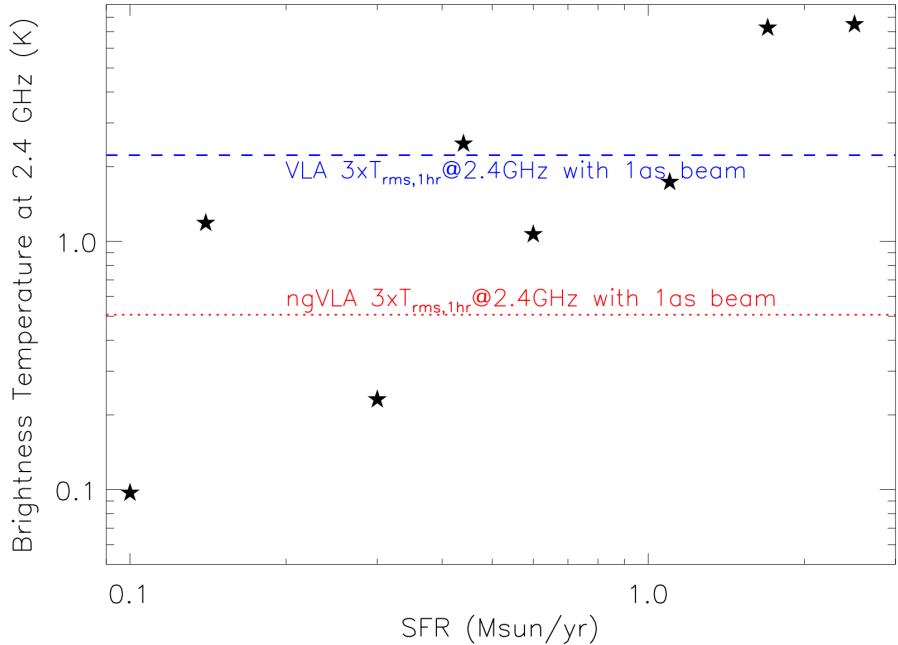}
\vspace{-0.5cm}
\caption{\footnotesize{\it Predictions for the radio brightness temperature of NS-NS and BH-NS binary merger host galaxies located at 120\,Mpc. These predictions are based on our current knowledge of short GRB hosts (see Table 1). The ngVLA will reach a surface brightness sensitivity of $\approx 0.1688$\,K in 1 hr with $\approx 1''$ resolution at 2.4 GHz, thus resolving all of these hosts at $d_L\lesssim 120$\,Mpc (red dotted line). More than half of the short GRB-like host galaxies at $d_L\lesssim 120$\,Mpc are inaccessible to the current Jansky VLA in 1\,hr (blue dashed line)}.}
\end{center}
\end{figure}

Radio continuum emission from galaxies traces the SF rate (SFR) during the last $\sim 50-100$\,Myr \citep[e.g.,][]{Murphy2011}, and could thus help track the connection between mergers and recent SFR rate. In fact, due to the age-dust-metallicity degeneracy, the optical/UV emission alone is an unreliable measure of SFR in dusty galaxies. Of course, radio measurements are subject to AGN contamination.  That said, radio measurements have been demonstrated to be an important complementary tool to constrain the SFR in the past, for example for short GRB hosts.  Short GRB mergers are believed to be cosmological compact binary mergers containing at least one NS, so are immediately pertinent comparisons. Radio observations of the host galaxies of GRB\,071227 and GRB\,120804A provided clear evidence for SFRs at least an order of magnitude larger than those derived using optical data alone \citep[][]{Nicuesa2014}, suggesting either a starburst origin or an AGN contribution. The host galaxy NGC4993 of the recently detected binary NS merger GW170817, associated with the short GRB\,170817A, shows a radio SFR of $\approx 0.1$\,M$_{\odot}$/yr, approximately $ 10\times$ higher than the one estimated using the galaxy broad-band photometry, indicative of an AGN dominating the radio emission \citep[][]{Blanchard2017}. We note that while the FIR is also an extinction free SFR tracer, ground-based extinction-free radio observations are advantageous compared to satellite-based FIR observations. 

The study of sub-galactic environments (achievable via \textit{resolved} multi-wavelength studies of the host galaxies) can be used to gain other clues to the progenitors of compact binary mergers, such as the presence or absence of a spatial association with star formation or stellar mass, and the distribution of offsets of EM counterparts with respect to their host galaxy light \citep[which can be used to map natal kicks, e.g.][ and references therein]{Belczynski2006,Behroozi2014}. 

While currently most of the cosmological short GRB host galaxies remain undetected in the radio, a sensitive array like the ngVLA will be able to resolve short GRB-like hosts within the LIGO horizon distance.  
To show this, in Table 1 we collected the (optical/UV) SFR and optical sizes  of short GRB hosts available in the literature \citep{Berger2014}. We use the measured SFR to calculate the expected  host galaxy luminosity at 1.4\,GHz via the \citet{Murphy2011} SFR-to-non-thermal radio emission calibration relation: $\left(\frac{\rm SFR_{1.4\rm\,GHz}}{M_{\odot}\rm yr^{-1}}\right)=6.35\times10^{-29}\left(\frac{L_{1.4\rm\,GHz}}{\rm erg\,s^{-1}Hz^{-1}}\right).$
To estimate the corresponding luminosity at $\approx 2.4$\,GHz (the central frequency of the ngVLA lowest frequency band), we extrapolate from the 1.4\,GHz luminosity using a spectral index of $\approx -0.75$ (as appropriate for non-thermal galaxy emission).  
Since radio emission from local galaxies is observed to be more concentrated than in the optical \citep{Condon2002,Murphy2017}, we then use the effective galaxy radius $r_e$ from optical/IR observations as FWHM size of the galaxy at radio wavelengths \citep{Fong2013}, and calculate the brightness temperature as:
$T=1.36\frac{\lambda^2}{\theta^2}S$,
where $\lambda$ is the observed wavelength in cm, $\theta$ is the FWHM radio size in arcsec, and $S$ is the flux density in mJy at the considered wavelength ($\lambda \approx 12.5$\,cm for observations at 2.4\,GHz). Our results are plotted in Fig. 4. The ngVLA will reach a surface brightness sensitivity of $\approx 0.1688$\,K in 1 hr with $\approx 1''$ resolution at 2.4 GHz, thus resolving most of these hosts at $d_L\lesssim 120$\,Mpc. The current Jansky VLA in its A configuration would offer a comparable resolution of $\approx 1''$ at 2.4\,GHz, but with a worse surface brightness sensitivity in 1\,hr of $\approx 0.742$\,K. More than half of the host galaxies in Fig. 4 have a brightness temperature $\lesssim 3\times 0.742\,{\rm K}\approx 2.2$\,K, thus the ngVLA is likely to greatly enlarge the sample of resolvable short GRB-like host galaxies.

\section{Summary and conclusion}
\label{sec:5}
We have investigated three new scientific opportunities that would emerge in time-domain astrophysics if a facility like the ngVLA were to work in tandem with ground-based GW detectors, and shown that:
\begin{enumerate}
 \item The ngVLA, thanks to its superior sensitivity, will substantially extend the reach of the current Jansky VLA for discovering and characterizing radio afterglows of NS-NS/BH-NS mergers, and their linear polarization fractions, to distances $3\times$ as large (for which event rates are a factor of $\sim 30\times$ larger);
  \item The ngVLA+VLBA, thanks to its superior sensitivity at the longest baselines, can enable direct size measurements of radio afterglows from NS-NS and/or BH-NS mergers, probe directly the dynamics of merger ejecta and, when VLBI is paired with linear polarization studies, constrain the magnetic field structure;
 \item The ngVLA, thanks to its improved surface brightness sensitivity at lower frequencies, can also enable resolved studies of NS-NS and/or BH-NS mergers host galaxies within distances of $\approx 120$\,Mpc. \end{enumerate}

\acknowledgements \footnotesize{A.C. thanks Chris Carilli and Remy Indebetouw for helping with ngVLA simulations in CASA. A.C. and D.C. acknowledge support form NSF CAREER award \#1455090. The NRAO is a facility of the National Science Foundation operated under cooperative agreement by Associated Universities, Inc.}

\begin{footnotesize}
\bibliographystyle{aasjournal}
\bibliography{ngVLA_Corsi_2018Oct15}
\end{footnotesize}

\end{document}